\documentclass[aps,prl,twocolumn,preprintnumbers,floatfix]{revtex4}
\usepackage[dvips]{graphics}
\usepackage{amsmath,amsfonts,bm}

\begin{document}

\newcommand\<{\langle}
\renewcommand\>{\rangle}
\renewcommand\d{\partial}
\newcommand\LambdaQCD{\Lambda_{\textrm{QCD}}}
\newcommand\tr{\mathop{\mathrm{Tr}}}
\newcommand\+{\dagger}
\newcommand\g{g_5}
\newcommand\aone{a}
\newcommand\mev{\rm MeV}
\newcommand\gev{\rm GeV}

\preprint{EFI-07-08}

\affiliation{Enrico Fermi Institute and Department of Physics, University of Chicago,
Chicago Illinois 60637, USA}
\title{Baryon Number-Induced Chern-Simons Couplings of Vector and Axial-Vector Mesons  in Holographic QCD}
\author{Sophia K. Domokos}
\affiliation{Enrico Fermi Institute and Department of Physics, University of Chicago,
Chicago Illinois 60637, USA}
\author{Jeffrey A. Harvey}
\affiliation{Enrico Fermi Institute and Department of Physics, University of Chicago,
Chicago Illinois 60637, USA}

\date{April 2007}

\newcommand\sect[1]{\emph{#1}---}

\begin{abstract}
We show that holographic models of QCD predict the presence of a Chern-Simons coupling between vector and axial-vector mesons at finite baryon density. In the AdS/CFT dictionary, the coefficient of this coupling
is proportional to the baryon number density, and is fixed uniquely in the five-dimensional holographic
dual by anomalies in the flavor currents. 
For the lightest mesons, the coupling mixes transverse $\rho$
and $a_1$ polarization states.
At sufficiently large baryon number densities, it produces an
instability, which causes the $\rho$ and $a_1$ mesons to condense 
in a state breaking both rotational
and translational invariance. 
\end{abstract}
\keywords{QCD, AdS-CFT Correspondence}
\pacs{11.25.Tq, 
11.10.Kk, 
11.25.Wx, 
}
\maketitle

\section{\label{sec:level1}Introduction}
Models which use the gravity/gauge correspondence to treat strongly-coupled QCD as a five-dimensional theory of gravity have progressed dramatically in recent years \cite{Maldacena:1997re,Gubser:1998bc,Witten:1998qj}. 
Particularly at high energies, these theories differ significantly from QCD -- yet those models which incorporate light quarks \cite{Karch:2002sh} and chiral symmetry-breaking of the form observed in QCD \cite{Sakai:2004cn}
do capture much of the important low-energy structure of the
theory, and give rise to a spectrum of mesons whose masses, decay constants, and couplings
match those of QCD to within 20\%. 

The gravity/gauge approach includes both top-down models of QCD arising from $D$-brane constructions in string theory \cite{Sakai:2004cn}, and bottom-up phenomenological
models, which attempt to capture the essential dynamics using a simple choice of five-dimensional
metric ($AdS_5$) and a minimal field content consisting of a scalar $X$ and gauge fields $A_{L\mu}^a$,  and $A_{R \mu}^a$ \cite{Erlich:2005qh,DaRold:2005zs}. These fields are holographically dual
to the quark bilinear $\bar q^\alpha q^\beta$, and to the $SU(N_f)_L \times SU(N_f)_R$ flavor currents
$\bar q_L \gamma^\mu t^a q_L$ and $\bar q_R \gamma^\mu t^a q_R$ of QCD, respectively.

These holographic models can be used to study QCD at finite baryon density
\cite{Horigome:2006xu,Kobayashi:2006sb}. In this paper
we focus on a novel effect, in which a Chern-Simons term leads to
mixing between vector and axial-vector mesons.
We will use the model introduced in~
\cite{Erlich:2005qh,DaRold:2005zs} and for the most part follow the conventions
and notation of \cite{Erlich:2005qh}. 

\section{The Model}
We work in a slice of $AdS_5$ with metric
\begin{equation}\label{onea}
ds^2 = \frac{1}{z^2} \left( -dz^2 + dx^\mu dx_\mu \right), \qquad 0 < z \le z_m ~.
\end{equation}
The fifth coordinate, $z$, is dual to the energy scale of QCD. We generate confinement
by imposing an IR cutoff $z_m$, and specifying the IR boundary conditions on the fields.
The UV behavior, meanwhile, is governed by $z \rightarrow 0$.

In AdS/CFT calculations, boundary contributions to the action must be treated with care. In 
the full AdS space, the only boundary is in the UV (at $z=0$). UV-divergent contributions to the action
and to other quantities are canceled by counterterms. 
For details see \cite{Bianchi:2001kw,Karch:2005ms}. In the model at hand, 
the IR boundary at $z=z_m$ may contribute to the action. We follow 
the approach of \cite{Erlich:2005qh,DaRold:2005zs} by (1) dropping IR boundary terms, and (2) taking 
parameters normally fixed by IR boundary conditions on the classical solution as input parameters of the model.

We generalize the gauge symmetry to $U(N_f)_L \times U(N_f)_R$ and add a Chern-Simons
term which gives the correct holographic description of the QCD flavor anomalies \cite{Witten:1998qj}. 
The Chern-Simons term does not depend on the metric and on general grounds will be present
in any holographic dual description of QCD.  The $U(1)$ axial symmetry in QCD is anomalous,
but in the spirit of the large $N_c$ approximation we treat it as an exact symmetry of QCD with
massless quarks. Including the anomaly would not affect our conclusions. 

The Lagrangian is thus
\begin{equation}\label{oneb}
S = \int d^4xdz \sqrt{g} \tr \left[ |DX|^2 + 3|X|^2 - \frac{1}{4g_5^2} (F_L^2 + F_R^2) \right] + S_{CS}~.
\end{equation}
The Chern-Simons term is given by
\begin{equation}\label{csterm}
S_{CS}= \frac{N_c}{24 \pi^2} \int \left[ \omega_5(A_L) - \omega_5(A_R)  \right] 
\end{equation}
where $d \omega_5 = \tr F^3$,
$N_c$ is the number of colors, and $A_{L,R} = \hat A_{L,R} \hat t + A_{L,R}^a t^a$ where $t^a$ are
the generators of $SU(N_f)_{L,R}$ normalized so that $\tr t^a t^b = \delta^{ab}/2$ and
$\hat t = {\bf 1}/\sqrt{2 N_f}$ is the generator of the $U(1)$ subalgebra  of $U(N_f)$. 
In what
follows, we take $N_f=2$ so that $a=1,2,3$. We will often work with the vector and axial-vector fields
$V=(A_L+A_R)/2$ and $A= (A_L-A_R)/2$.

\section{Classical Background}
We expand around a nontrivial solution to the classical equations of motion for the scalar
$X$.
Following \cite{Erlich:2005qh,DaRold:2005zs} we find the scalar background
\begin{equation}\label{twoa}
X_0(z) = \left( \frac{1}{2} M z + \frac{1}{2}\Sigma z^3\right)\equiv \frac{v(z)}{2}{\bf 1} 
\end{equation}
where the coefficient $M$ of the non-normalizable term is proportional to the
quark mass matrix, and $\Sigma$ is the $\bar q q$ expectation value. We take both $M$ and $\Sigma$ to be diagonal: $M \equiv m_q {\bf 1}$ and 
$\Sigma \equiv \sigma {\bf 1}$. 
As shown in \cite{Erlich:2005qh,DaRold:2005zs}, we can fix the five-dimensional coupling
$g_5$ by comparison with the vector current two-point function in QCD at large
Euclidean momentum. This leads to the identification
\begin{equation}\label{twoaa}
g_5^2 = \frac{12 \pi^2}{N_c}~.
\end{equation}
The model is thus  defined by three parameters: $z_m$, $m_q$ and $\sigma$.
Note that including the $U(1)$ gauge fields and Chern-Simons coupling
does not mandate the addition of any new parameters. We use $z_m = 1/(346 ~\mev)$,
$m_q=2.3~ \mev$ and $\sigma = (308 ~\mev)^3$, which correspond to values found through
a global fit to seven observables (Model B) in \cite{Erlich:2005qh}. 

A background with a static, constant quark density is described by a classical solution
to the equation of motion for the time component of the $U(1)$ vector gauge field $\hat V_\mu$, which
is dual to the quark
number current.  Solving the  
$\hat V_0$ equation of motion at zero four-momentum yields
\begin{equation}\label{twob}
\hat V_0(z) = A + \frac{1}{2} B z^2 ~.
\end{equation}
By the general philosophy of AdS/CFT, the coefficient of the non-normalizable term, $A$,
is proportional to the coefficient with which the operator dual to $\hat V_0$
enters the gauge theory Lagrangian. Since $\hat V_\mu$ is dual to the quark number current,
$A$ must be proportional to the quark chemical potential. Meanwhile, the coefficient of 
the normalizable term, $B$, is proportional to
the expectation value of the operator dual to $\hat V_0$: the quark number density.
We now obtain the normalizations of $A$ and $B$. The action evaluated for the
background Eq.\ \eqref{twob} is given by a boundary term:
\begin{equation}\label{twoc}
S = \frac{1}{2 g_5^2} \int d^4 x \frac{1}{z} \hat V_0 \partial_z \hat V_0 |_{z=0} = 
\frac{1}{2 g_5^2} AB \int d^4 x ~.
\end{equation}
At finite temperature and baryon number, the Euclidean action is equal to the grand canonical
potential. Using Eq.\ \eqref{twoaa}, this implies that
\begin{equation}\label{twod}
A B = \frac{24 \pi^2}{N_c} n_q \mu_q
\end{equation}
with $n_q $ the quark number density and $\mu_q$ the quark chemical potential. 
To fix $A$ we separate $U(N_f)_{L,R}$ into
$U(1)_{L,R}$ and $SU(N_f)_{L,R}$ components and note that the Chern-Simons term contains
the coupling
\begin{equation}\label{twoe}
\frac{N_c}{24 \pi^2} \frac{3}{8} \int d^4 x dz \epsilon^{MNPQ} ( \hat A_{0}^L \tr F_{MN}^L F_{PQ}^L -
\hat A_0^R \tr F^R_{MN} F^R_{PQ} )
\end{equation}
where the indices $M,N,P,Q$ run over $1,2,3,z$ and the trace is over $SU(N_f)$. Defining the $SU(N_f)_{L,R}$
instanton numbers by
\begin{equation}\label{twog}
n_{L,R} = \frac{1}{32 \pi^2} \int d^3x dz \epsilon^{MNPQ} \tr F^{L,R}_{MN} F^{L,R}_{PQ}
\end{equation}\label{twoh}
and taking $\hat A^{L,R}_0$ constant, this reduces to the coupling
\begin{equation}\label{twoi}
\frac{N_c}{2} \int dx^0 \left( \hat A^L_0 n_L - \hat A^R_0 n_R \right) ~.
\end{equation}
Using the connection between instantons and Skyrmion configurations of the pion field
carrying non-zero baryon number \cite{Atiyah:1989dq,Son:2003et,Nawa:2006gv,Hong:2007kx,Hata:2007mb}, we can interpret an instanton
with $n_L = -n_R = N_b$ as a state with baryon number $N_b$.
Eq.\ \eqref{twoi} then fixes $A= \mu_b/N_c = \mu_q$ with $\mu_q$ the quark
chemical potential; Eq.\ \eqref{twod} fixes $B=24 \pi^2 n_q/N_c$.

\section{Quadratic Action}
In vacuum, the spectrum of the
theory consists of towers of scalar, vector,  pseudoscalar, and axial-vector mesons given
by mode-expanding the five-dimensional fields along the holographic ($z$) direction, and integrating over $z$. 
In this section, we identify the spectrum of excitations and their dispersion relations at non-zero baryon density by expanding the action to quadratic order  around the background given by
Eqns.\ \eqref{twoa},\eqref{twob}.

We focus on the $\pi$ mesons and the isospin triplet vector $\rho$ and axial-vector $a_1$ mesons, ignoring contributions from heavier mesons, and from the scalar $\sigma$ which arises from fluctuations in
the magnitude of $X$. Couplings similar to those for the $\rho-a_1$ mesons exist for the isoscalar $\omega$ and $f_1$ mesons.
For simplicity, we omit these as well.

Pions arise
as Nambu-Goldstone modes associated with the breaking of $U(N_f)_L \times U(N_f)_R$
to $U(N_f)_V$.  We write $X(x,z) = X_0(z) \exp(i 2 \pi^a t^a)$ and expand to quadratic order in $\pi^a$. The four-dimensional pion field is obtained by writing $\pi^a(x,z) = \pi^a(x)
\psi_{\pi}(z)$. Similarly, the $\rho^a$ and $a_1$ mesons appear by writing 
$V_\mu^a(x,z) =  g_5 \rho^a_\mu(x) \psi_\rho(z)$, 
$A_\mu^a(x,z) =  g_5 \aone^a_\mu(x) \psi_{\aone}(z)$.
The wave functions $\psi_{\pi}(z)$, $\psi_\rho(z)$,
and $\psi_{\aone}(z)$ are solutions of the quadratic equations of motion for fields with four-momentum
$q^2=m^2$ and with boundary conditions $\psi(0)=\partial_z \psi(z_m)=0$.  For details see 
\cite{Erlich:2005qh,DaRold:2005zs}.

Making the above substitutions and expanding to quadratic order yields the four-dimensional action
\begin{eqnarray}\label{foura}
S = \int   d^4x ~  \Biggl[\frac{1}{2} \partial_\mu \pi^a \partial^\mu \pi^a 
- \frac{1}{2} m_\pi^2 \pi^a \pi^a 
- \frac{1}{4}(\rho^a_{\mu \nu} )^2  \nonumber \\ 
- \frac{1}{4} ( \aone^a_{\mu \nu} )^2  +
 \frac{1}{2} m_\rho^2 \rho^a_\mu \rho^{a \mu}
+ \frac{1}{2} m_{\aone}^2 \aone^a_\mu \aone^{a \mu} \nonumber \\
+ \mu \epsilon^{ijk}\left( \rho^a_i \partial_j \aone^a_k + \aone^a_i \partial_j \rho^a_k \right)\Biggr]~,
\end{eqnarray}
with $\rho_{\mu \nu},a_{\mu \nu}$ the field strengths for $\rho_\mu, a_\mu$. The Chern-Simons term with coefficient $\mu$ mixes the $\rho$ and $a_1$ mesons. It
arises from reduction of a term of the form $\int d \hat V \tr A dV$ in the expansion of Eq.\ \eqref{csterm}.

As usual, to obtain Eq.\ \eqref{foura} one must remove the mixing between $a^a_\mu$ and
$\partial_\mu \pi^a$ by performing the transformation $a_\mu^a \rightarrow a_\mu^a
 + \xi \partial_\mu \pi^a$
and then rescaling the pion field to obtain a canonical kinetic energy term \cite{Gasiorowicz:1969kn}.
This leads to a pion contribution to the Chern-Simons term. A total spatial derivative, it does not contribute to the equations of motion and may be
dropped. 

Since the $\rho $ has $J^{PC}= 1^{--}$ and the $a_1$ has $J^{PC}=1^{++}$, the Chern-Simons coupling is even under $P$ and odd under $C$. This is indeed consistent with a background having non-zero baryon number, which preserves $P$ and violates $C$: the coupling is rotationally invariant, but not Lorentz invariant due to the preferred rest frame of the baryons.

We can deduce the existence of the Chern-Simons coupling in four-dimensional terms as follows.
The reduction of the five-dimensional Chern-Simons term \cite{Sakai:2004cn,Hill:2006wu} gives rise to the usual gauged
WZW action \cite{Wess:1967jq,Witten:1983tw,Kaymakcalan:1983qq},
as well as a set of couplings which arise from inexact bulk terms. These include
a $\rho-a_1- \omega$ coupling which, in the presence of a coherent $\omega$ field in nuclear
matter, gives rise to a coupling of the form given in Eq.\ \eqref{foura}.
The $\rho-a_1-\omega$ coupling has been considered previously in a general discussion of
chiral effective Lagrangians \cite{Kaiser:1990yf}, and is implicit in the formulae of
\cite{Sakai:2005yt}.  Related terms appear in \cite{Hill:2006ei,Hill:2006wu}. 
In AdS/QCD, different forms of the gauged WZW action can be obtained by the addition of UV counterterms \cite{Panico:2007qd}, but these
will not cancel the Chern-Simons coupling and lead to explicit breaking of
chiral symmetry beyond that given by the quark mass term in Eq.\ \eqref{twoa}.

The mass of the $\rho$ meson is given by $m_\rho = 2.405/z_m$, while $m_{a_1}$ must be
determined from a numerical solution of the equation of motion.  Model B of \cite{Erlich:2005qh}
finds $m_\rho= 832 ~ \mev$, $m_{a_1} = 1200 ~ \mev$ which should be compared to the
experimental values $m_\rho = 775.8 \pm 0.5 ~ \mev$ and $m_{a_1} = 1230 \pm 40 ~ \mev$
\cite{Eidelman:2004wy}.
The parameter $\mu$ in the Chern-Simons coupling is given by
\begin{equation}\label{fourb}
\mu = 18 \pi^2 n_b z_m^2 I
\end{equation}
where $I$ is the dimensionless overlap integral
\begin{equation}
I = \frac{1}{z_m^2} \int_0^{z_m} dz z \psi_\rho(z) \psi_{a_1}(z) ~.
\end{equation}
Numerical evaluation of the integral gives $I = 0.54$. 
A typical baryon density in nuclear matter, $n_b^0 \simeq 0.16/({\rm fermi})^3$ , gives
\begin{equation}\label{muval}
\mu \simeq  1.05 ~ \gev \left( \frac{n_b}{n_b^0} \right) ~.
\end{equation}
                                          
\section{Phenomenological applications}
We now outline two potentially observable consequences of the Chern-Simons 
coupling between the $\rho$ and $a_1$. Details will appear
elsewhere.

\subsection{\label{sec:level2}Mixing of transverse $\rho$ and $a_1$ states}
We consider plane-wave solutions to the equations of motion resulting from Eq.\ \eqref{foura},
dropping the pion fields and focusing on the $\rho$ and $a_1$ dispersion relation
and polarization vectors. Without loss of generality, we consider
propagation along $x^3$:
\begin{equation}\label{sixa}
\rho_\mu(x)  = \epsilon^\rho_\mu(q) e^{-i q \cdot x}, \qquad
                                          a_\mu(x)  = \epsilon^a_\mu(q) e^{-iq \cdot x} 
\end{equation}  
with $q=(q_0,0,0,q_3)$. For convenience, we suppress the $SU(2)$ indices in the following.
The components $\rho_0$, $\rho_3$, $a_0$, and $a_3$ have standard dispersion relations, unaffected by the Chern-Simons coupling. The transverse components $\rho_1$, $\rho_2$, $a_1$, and $a_2$
mix through a derivative coupling. The equations of motion yield the
dispersion relation for the transverse polarizations
\begin{equation}\label{disp}
q_0^2 - q_3^2 = \frac{1}{2}(m_\rho^2 + m_{a_1}^2) \pm \frac{1}{2} \sqrt{(m_{a_1}^2 - m_\rho^2)^2 +
16 \mu^2 q_3^2} ~.
\end{equation}
The lower sign in Eq.\ \eqref{disp} gives a state which is pure $\rho$ as $q_3 \rightarrow 0$.
At non-zero $q_3$, it is a mixture of transverse $\rho$ and $a_1$ states with orthogonal
polarization vectors:
\begin{equation}\label{rhopol}
\epsilon^a_1 =  \frac{i {\cal M}^2(q_3)}{2 \mu q_3} \epsilon^\rho_2, \qquad
\epsilon^a_2 = - \frac{i {\cal M}^2(q_3) }{2 \mu q_3}  \epsilon^\rho_1
\end{equation}
where we have defined $\Delta^2 = m_{a_1}^2 - m_\rho^2$ and
${\cal M}^2(q_3) = (\sqrt{\Delta^4 + 16 \mu^2 q_3^2} - \Delta^2)/2 $.  
The upper sign in Eq.\ \eqref{disp} gives a pure $a_1$ state for $q_3 =0$, while for non-zero $q_3$, 
 \begin{equation}\label{apol}
\epsilon^\rho_1 = - \frac{i {\cal M}^2(q_3)}{2 \mu q_3}  \epsilon^a_2, \qquad
\epsilon^\rho_2 = \frac{i {\cal M}^2(q_3)}{2 \mu q_3}  \epsilon^a_1 ~.
\end{equation}  

For $\mu$ greater than some momentum-dependent critical value, the dispersion relation 
Eq.\ \eqref{disp} leads to tachyonic modes (modes having $dq_0/dq_3>1$). For very large momenta,
this critical value becomes 
\begin{equation}\label{mucrit}
\mu_{\rm crit} = \sqrt{(m_\rho^2 + m_{a_1}^2)/2} \simeq 1.09 ~ \gev ~.
\end{equation}
For a range of $\mu$ below $ \mu_{\rm crit}$ the dispersion relation with the lower sign in
Eq.\ \eqref{disp} exhibits interesting anomalous behavior, the analysis of which is beyond the
scope of this letter.

It would be interesting to explore signatures of these mixed polarization states
in the quark-gluon plasma and in nuclear matter.
                                
\subsection{Vector Meson Condensation}

To identify the tachyonic instability which occurs for $\mu>\mu_{\rm crit}$ we
start with the energy density corresponding to Eq.\ \eqref{foura} for the diagonal component of the $\rho$ and $a$ fields, $a^a= a \delta^{a3}$,
$\rho^a = \rho \delta^{a3}$. Completing the square and dropping the terms involving the electric
components of the field strengths, which play no role in the instability, we find
\begin{eqnarray}\label{enden}
{\cal H} &=&  \frac{1}{2} (m_a^2 - \mu^2) \vec a \cdot \vec a  + \frac{1}{2} (m_\rho^2 - \mu^2) \vec \rho \cdot \vec \rho \nonumber \\
&+& \frac{1}{2}( \vec B_a - \mu \vec \rho)^2 + \frac{1}{2} ( \vec B_\rho - \mu \vec a)^2   
\end{eqnarray}
where $\vec B_\rho = \vec \nabla \times \vec \rho$ , $\vec B_a = \vec \nabla \times \vec a$.

Applying the ansatz
\begin{equation}\label{ansatz}
\vec a = v \cos(\mu x_3) \hat x_2, \qquad \vec \rho = v \sin(\mu x_3) \hat x_1~,
\end{equation}
the last two terms in Eq.\ \eqref{enden} vanish, while the average of the first two terms over $x_3$ is negative
for $\mu^2 > \mu^2_{\rm crit}$, leading to an instability to $v \ne 0$. 
Understanding the stabilization of the
configuration Eq.\ \eqref{ansatz} requires generalizing ${\cal H}$ to include higher order terms. Note that 
Eq.\ \eqref{ansatz} breaks both rotational and translational symmetry, exhibiting a structure
similar to the smectic phase of liquid crystals which includes an interesting set of topological defects.

The critical value Eq.\ \eqref{mucrit} is remarkably close to the estimate
Eq.\ \eqref{muval} for $\mu$ at ordinary nuclear densities.  If this model is accurate then there should
be a condensate of vector and axial-vector mesons in nuclear matter with baryon densities at or slightly above
$n_b^0$. In ordinary nuclei, there are finite size effects as well as other corrections to the $\rho$ and
$a_1$ interactions which will have to be included to determine whether this condensate occurs. Neutron stars 
are more likely to produce such a condensate, as they are thought to contain matter at a density somewhat greater than $n_b^0$. The interplay between this condensate and other
conjectured effects in nuclear matter, such as pion condensation and color superconductivity,
deserves further study.

\section{Acknowledgements}
We thank O. Lunin and J. Rosner for helpful conversations. JH thanks the Galileo Galilei Institute in
Arcetri, Florence for hospitality while this work was being completed. 
The work of SD and JH  was supported in part by NSF
Grant No. PHY-00506630 and NSF Grant 0529954. Any opinions,
findings, and conclusions or recommendations expressed in this
material are those of the authors and do not necessarily reflect the
views of the National Science Foundation.

\end{document}